\documentstyle[11pt,newpasp,twoside,epsf]{article}
\def\edcomment#1{\iffalse\marginpar{\raggedright\sl#1\/}\else\relax\fi}
\reversemarginpar

\begin{document}
\title{Microlensing and the Physics of Stellar Atmospheres}
 \author{Penny D. Sackett}
\affil{Kapteyn Institute, 9700 AV Groningen, The Netherlands}
\affil{Anglo-Australian Observatory, P.O. Box 296, Epping NSW 1710, Australia}

\begin{abstract}
The simple physics of microlensing provides a well-understood tool with which to probe 
the atmospheres of distant stars in the Galaxy and Local Group with 
high magnification and resolution.  Recent results in measuring stellar surface
structure through broad band photometry and spectroscopy of high amplification 
microlensing events are reviewed, with emphasis on the dramatic expectations 
for future contributions of microlensing to the field of stellar atmospheres. 
\end{abstract}

\section{Introduction}

The physics of microlensing is simple.  For most current   
applications, the principles of geometric optics combined with 
one relation (for the deflection angle) from General Relativity is all that 
is required.  For observed Galactic microlensing events, 
the distances between source, lens and observer are large compared to intralens 
distances, so that small angle approximations are valid.  Although it is possible 
that most lenses may be multiple, $\sim$90\% of observed Galactic microlensing 
light curves can be modeled as being due to a single point lens.  
Usually, though not always (cf., Albrow et al.~2000), binary lenses can be considered 
static throughout the duration of the event.  

The magnification gradient 
near caustics is large, producing a sharply peaked lensing ``beam'' that sweeps 
across the source due to the relative motion between the lens and 
the sight line to the source (Fig.~1).  
Furthermore, the combined magnification of the multiple 
microimages (which are too close to be resolved with current techniques) 
is a known function of source position that is always greater than 
unity, so that more flux is received from the source during the lensing event.
The net result is a well-understood astrophysical tool that can simultaneously 
deliver high resolution and high magnification of tiny background sources. 
In Galactic microlensing, these sources are stars at distances of a few to 
a few tens of kiloparsecs.

The great potential of microlensing for the study of  
stellar polarization (Simmons, Willis, \& Newsam 1995; Simmons, 
Newsam \& Willis 1995; Newsam et al.~1998; Gray 2000), stellar spots 
(Heyrovsk\'y \& Sasselov 2000; Bryce \& Hendry 2000), and motion in circumstellar 
envelopes (Ignace \& Hendry 1999) will not be treated here.  
Instead, the focus will be on how the composition of spherically-symmetric 
stellar atmospheres can be probed by microlensing.  

\section{Caustic Transits}

The angular radius $\theta_{\rm E}$ of a typical Einstein ring is about  
two orders of magnitude larger than the size $\theta_*$ 
of a typical Galactic source star (few $\mu$as), but the gradients in 
magnification that generate source resolution effects are appreciable 
only in regions near caustics.   
For a single point lens, the caustic is a single point coincident 
with the position of the lens on the sky that must directly transit the 
background source in order to create a sizable finite source effect. 
The probability of such a point transit is  
of order $\rho \equiv \theta_*/\theta_{\rm E} \approx \, $2\%.  
The amount of resolving power will depend on the dimensionless impact 
parameter $\beta$, the distance of the source center from the point 
caustic in units of $\theta_{\rm E}$.  The first clear point caustic transit 
was observed in event MACHO~95-BLG-30 (Alcock et al. 1997).

\begin{figure}
\plottwo{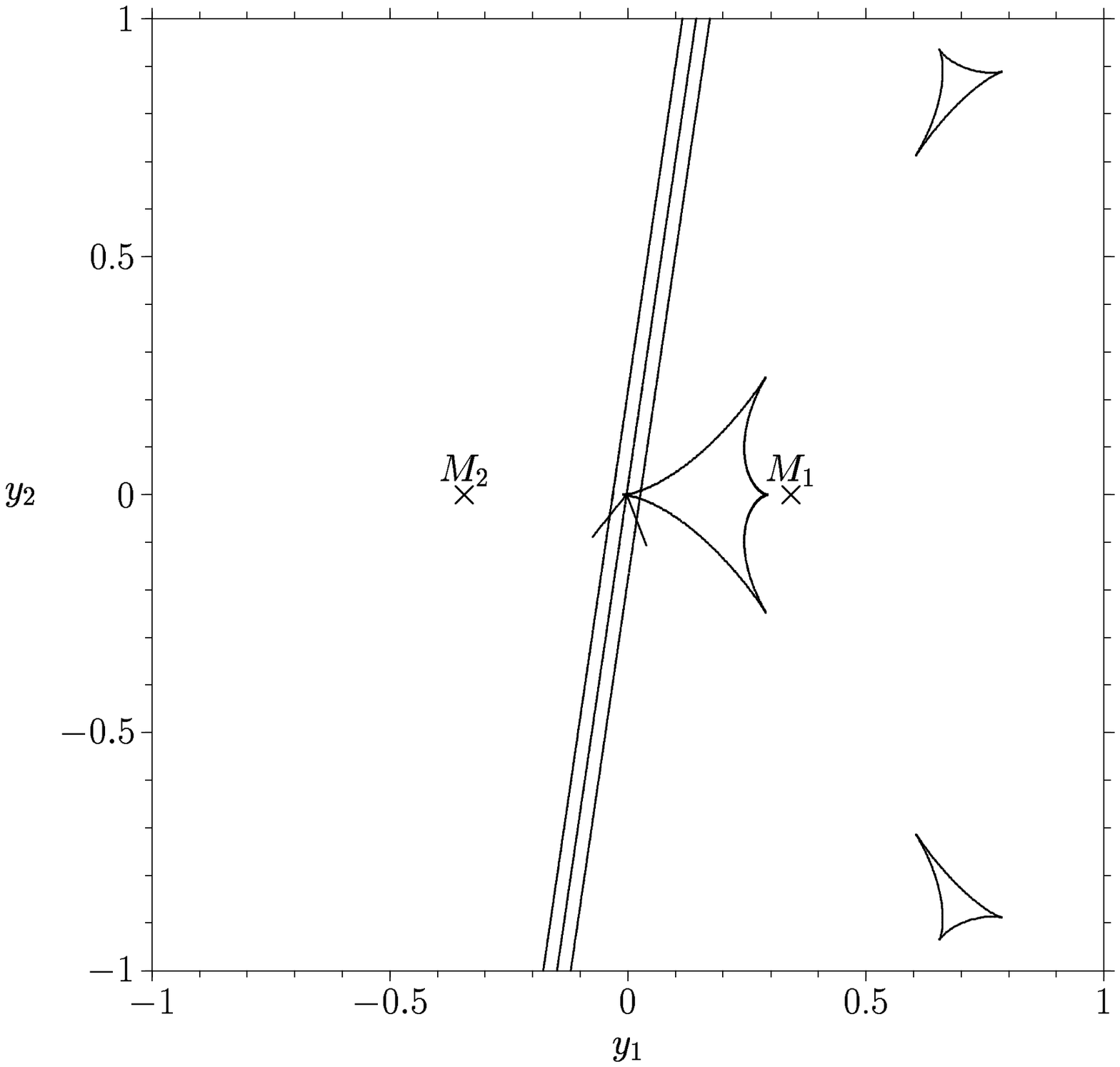}{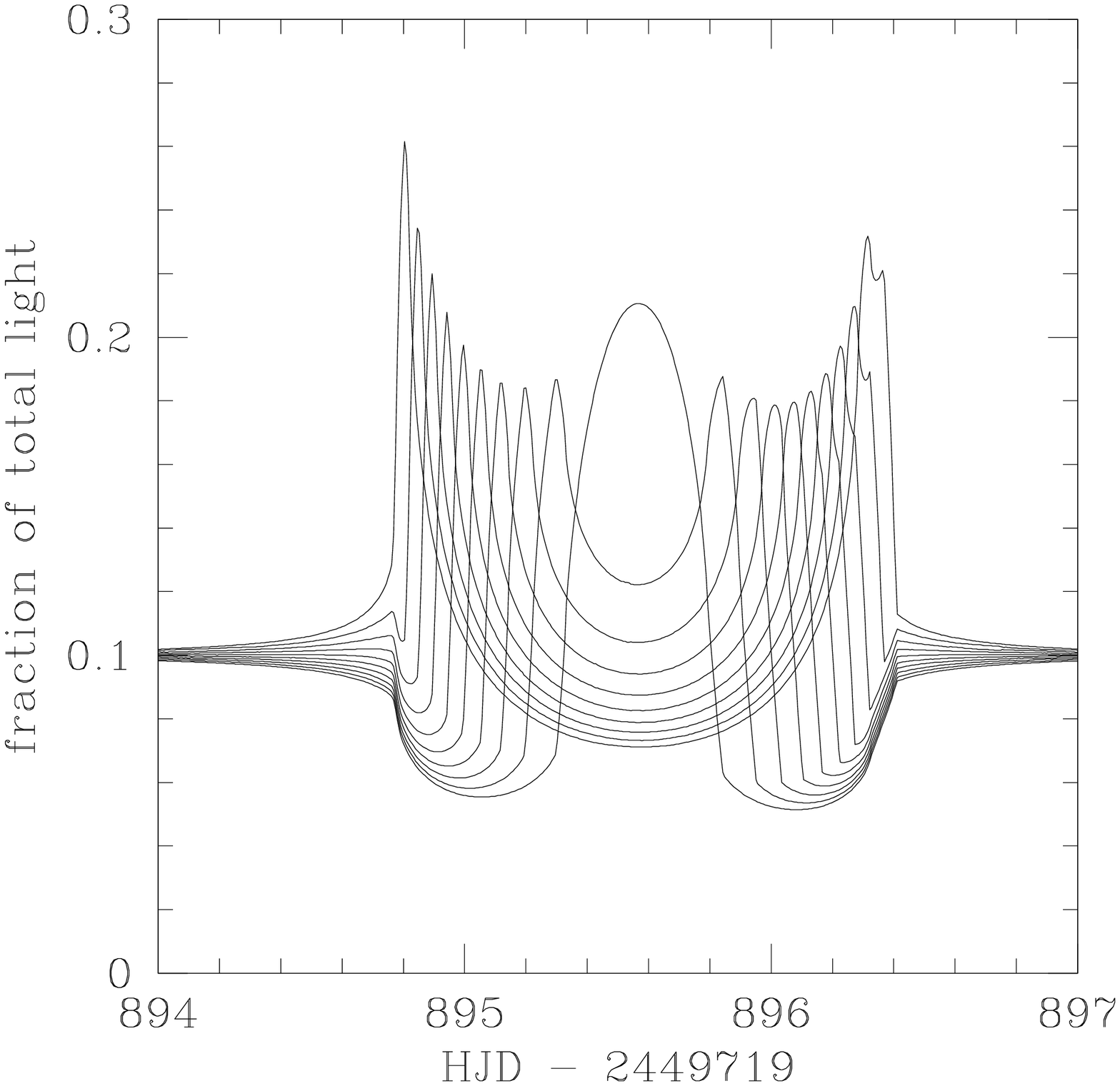}
\caption{{\bf Left:} The trajectory of the background source star 
passed over a caustic curve cusp of the binary lens MACHO~97-BLG-28. 
Units are $\theta_{\rm E}$. 
{\bf Right:} The fractional (magnified) flux 
in 10 concentric rings of equal area over the stellar disk are 
shown as a function of time during the two-day crossing (Albrow et al. 1999a).} 
\end{figure}

Lensing stellar binaries with mass ratios $0.1 \la q \equiv m_2/m_1 \la 1$ 
and separations $0.6 \la d \equiv \theta_{\rm sep}/\theta_{\rm E} \la 1.6$ 
generate extended caustic structures that cover a sizable fraction of the 
Einstein ring (see, eg., Gould 2000).  Since events generally are not alerted unless 
the source lies inside the Einstein ring, any alerted binary 
event with $q$ and $d$ in these ranges is highly likely to result 
in a caustic crossing.   If the source crosses the caustic 
at a position at which the derivative of the caustic curve is discontinuous, 
it is said to have been transited by a cusp.  For a given lensing binary, the 
probability of a cusp transit is of order $\rho \, N_{\rm cusps} \approx \, $10\%. 
Since $\sim$10\% of all events are observed to be lensing stellar binaries, 
the total cusp-transit probability is $\sim$1\%.  
To date, two cusp-crossing events have been 
observed, MACHO~97-BLG-28 (Albrow et al. 1999a) and MACHO~97-BLG-41 
(Albrow et al. 2000).  The remaining caustic crossing are transits of simple 
fold (line) caustics, which are observed in $\la$10\% of all events.  
Caustics thus present a non-negligible cross section to background 
stellar sources, with fold caustic transits being most likely by a factor of $\sim$5.

\begin{figure}
\vglue 2.2cm
\plotfiddle{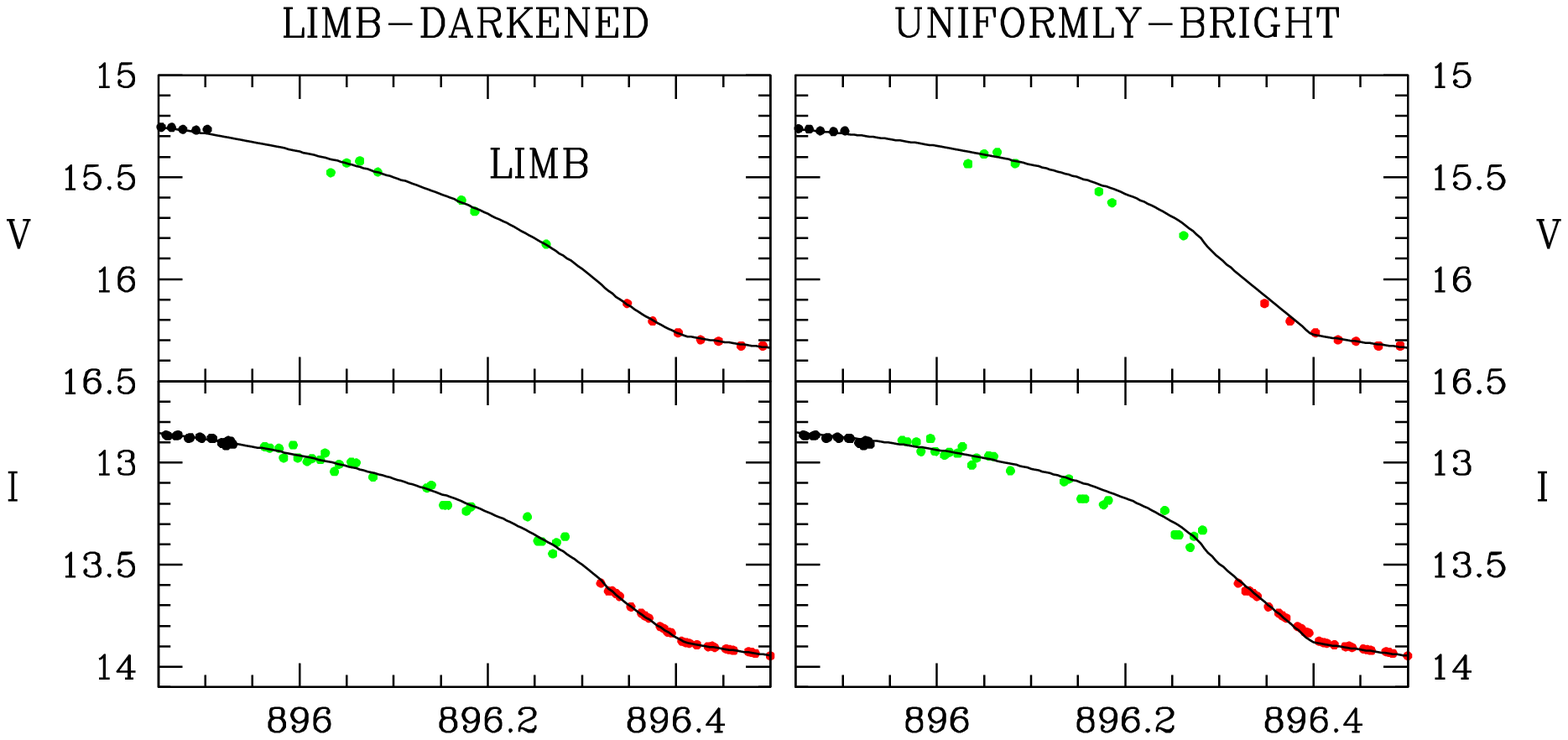} {4cm}{0}{55}{55}{-165}{-200}
\vglue 7.2cm
\plotfiddle{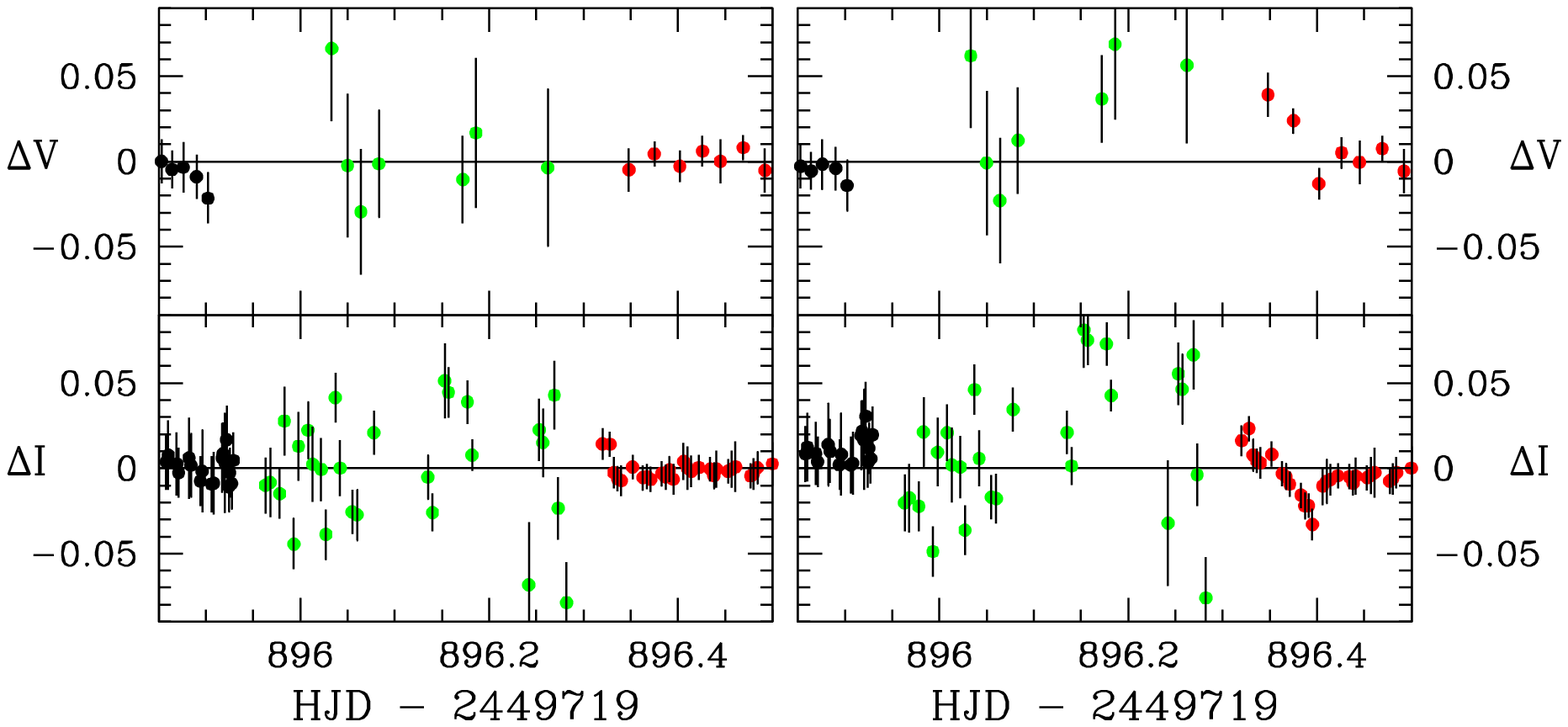} {4cm}{0}{55}{55}{-165}{0}
\vglue -8.3cm
\caption{{\bf Top:~}The $V$- and $I$-band light curves of MACHO~97-BLG-28 plummet 
as the trailing limb of the source exits the caustic. 
{\bf Bottom:} Residuals (in magnitudes) from the best models 
using a uniformly-bright (right) and
limb-darkened (left) stellar disk.  The same models are superposed 
on the light curves (Albrow et al. 1999a).  The varying scatter reflects 
different conditions at the three telescopes.
}
\vglue -5cm
\end{figure}
\vglue 4.5cm
The largest effect of a caustic crossing over an extended source 
is a broadening and diminishment of the light curve peak at transit 
that depends on the finite size ($\rho \neq 0$) of the source.  
If the angular size $\theta_*$ of the source star can be estimated independently 
(eg., from color-surface brightness relations), then the time required for the 
source to travel its own radius, and thus its proper motion $\mu$  
relative to the lens, can be determined from the light curve shape.  
Conversely, unless an independent method is available (see, Han 2000) 
to measure $\mu$ or $\theta_{\rm E}$, photometric 
microlensing cannot translate knowledge of the dimensional parameter 
$\rho$ into a measurement of source radius.  
What photometric or spectroscopic data alone {\it can\/} yield 
is a characterization of how the source profile 
differs from that of a uniform disk (Fig.~2).  Microlensing has already 
yielded such information for stars as distant as the 
Galactic Bulge and Small Magellanic Cloud.

\section{Recent Contributions of Microlensing to Stellar Physics}

The potential to recover profiles of stellar atmospheres 
from microlensing has been recognized for several 
years (Bogdanov \& Cherepashchuk 1995; Loeb \& Sasselov 1995; Valls-Gabaud 1995), 
but made possible only recently, due to the improved photometry and 
especially temporal sampling now obtained for a large number 
of events by worldwide monitoring networks.  
For only a few stars, most of which are supergiants or very nearby, has 
limb darkening been observationally determined by any technique.  
Microlensing has the advantage that: 
(1) many types of stars can be studied, including those quite distant;    
(2) the probe is decoupled from the source;  
(3) the signal is amplified (not eclipsed); and 
(4) intensive observations need only occur over one night.

\begin{figure}
\plotfiddle{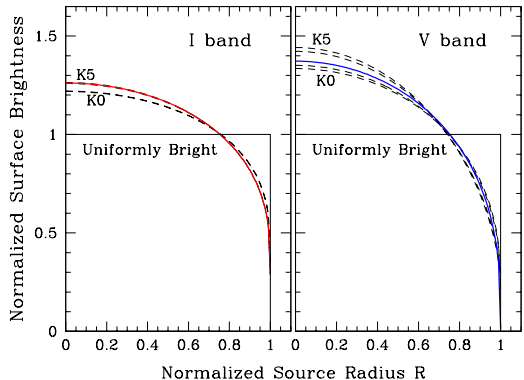}{4.0cm}{0}{35}{35}{-115}{-85}
\vglue 0.7cm
\caption{Stellar profiles deduced from microlensing light curve 
data (bold lines) and from atmospheric models (dashed lines) for the K-giant 
source star of MACHO~97-BLG-28 (Albrow et al. 1999a). 
} 
\end{figure}

\subsection{Limb Darkening}

The first cusp crossing was observed in MACHO 97-BLG-28, 
and led to the first limb-darkening measurement of a Galactic Bulge star 
(Albrow et al 1999a).
As the source crossed the caustic cusp, a characteristic anomalous 
bump was generated in the otherwise smooth light curve.  
First the leading limb, then the center, and finally the trailing limb 
of the stellar disk were differentially magnified (Fig.~1).  
Analysis of the light curve shape during the limb crossing 
allowed departures from a uniformly-bright stellar disk to be 
quantified (Fig.~2) and translated into a surface brightness profile in 
the $V$ and $I$ passbands.  A two-parameter limb-darkened 
model provided a marginally better fit than a linear model. 
Spectra provided an independent typing of the source as a KIII giant.  
The stellar profile reconstructed from the microlensing light curve 
alone is in good agreement with those from stellar atmosphere models (van Hamme 1993;  
Claret, Diaz-Cordoves, \& Gimenez 1995; Diaz-Cordoves, Claret, \& Gimenez) 
for K giants fitted to the same two-parameter (square-root) law (Fig.~3).

This first microlensing measurement of limb darkening was encouraging, 
but constructing realistic error bars for the results proved awkward.  
In traditional parameterizations for limb-darkening 
the coefficients $c_{\lambda}$ and $d_{\lambda}$ defined by 
\begin{equation}
I_{\lambda}(\theta) = I_{\lambda}(0) \, \left[ 1 - c_{\lambda} (1 - \cos \theta) 
- d_{\lambda} (1 - \cos^n \theta) \right]~~~~~{\rm where~} n = 0, 1/2, 2
\end{equation}
are correlated not only with each another, but also with other 
parameters in the microlensing fit because they carry information 
about the total flux $F$ of the source.   
(Here $\theta$ is the angle between the normal to the stellar surface 
and the line of sight.)  
A different parameterization was therefore constructed for the 
analysis of fold caustic crossings (Albrow et al. 1999b),  
\begin{equation}
I_{\lambda}(\theta) = \left< I_{\lambda} \right> \, \left[ 1 - \Gamma_{\lambda} 
(1 - {\frac{3}{2}} \, \cos \theta) \right]~~~~~{\rm where~} \left< I_{\lambda} \right>
\equiv {\frac{F}{\pi \theta_*^2}}~~~,
\end{equation}
which decouples the limb-darkening parameter $\Gamma_{\lambda}$ 
from the source flux.  
To first order, $c_{\lambda} = 3 \, \Gamma_{\lambda}/(\Gamma_{\lambda} + 2)$. 
This form was implemented in the analysis of the multiband data collected by five 
teams (Fig.~4) for the fold caustic crossing event MACHO~98-SMC-1 (Afonso et al. 2000).
The source star was typed from spectra to be an A-dwarf in the Small Magellanic Cloud 
(and thus a radius $\theta_* = 80$~nanoarcsec!).  
As expected, limb darkening decreases with increasing wavelength and at given 
wavelength is smaller for a hot dwarf than a cool giant (Figs.~3 \& 4). 
Unfortunately, no models of metal-poor A-dwarf stars were available for 
direct comparison with the 98-SMC-1 limb-darkening measurements.

\begin{figure}
\vskip -0.4cm
\plotfiddle{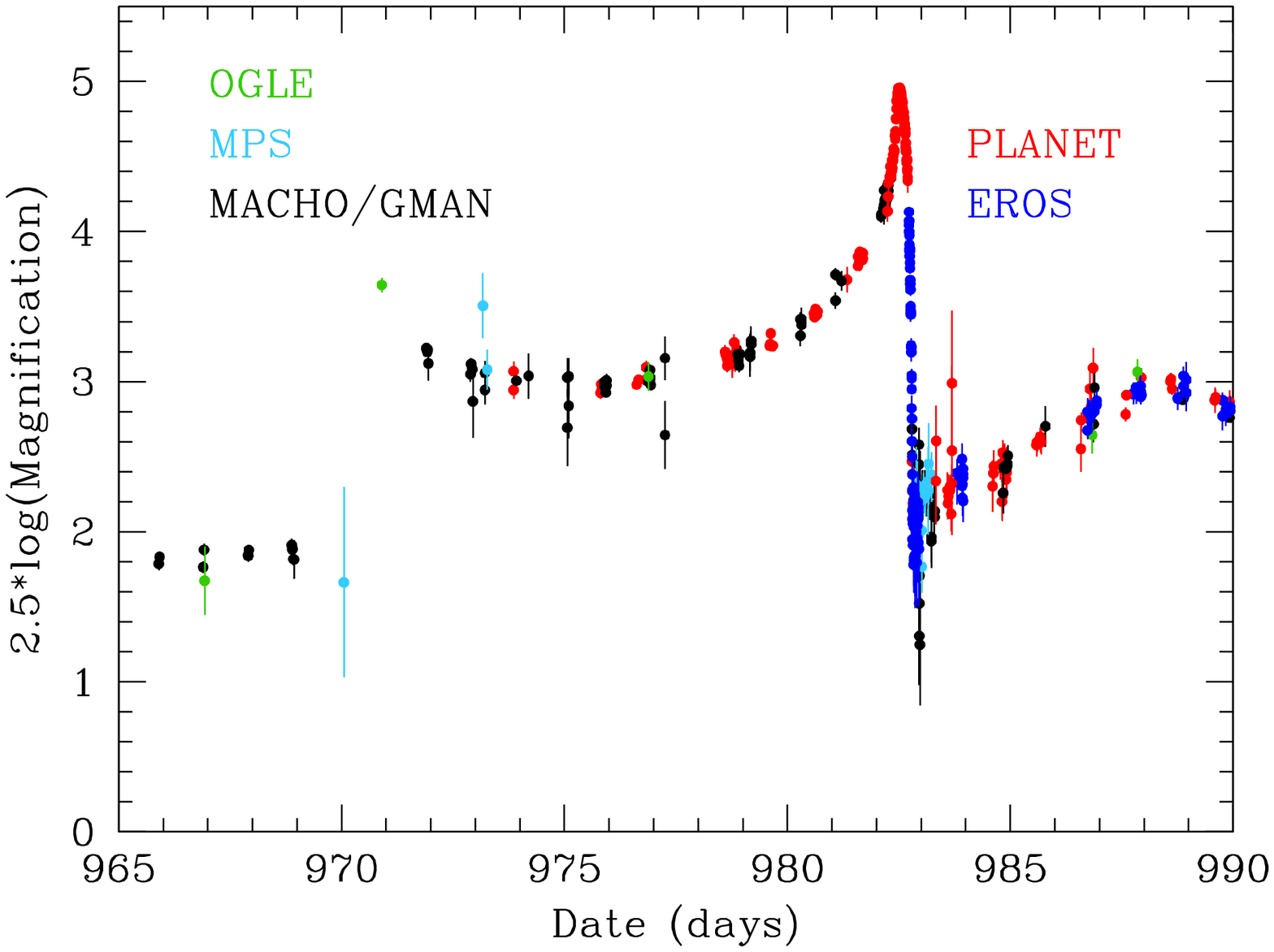}{4.0cm}{0}{35}{35}{-180}{-100}
\plotfiddle{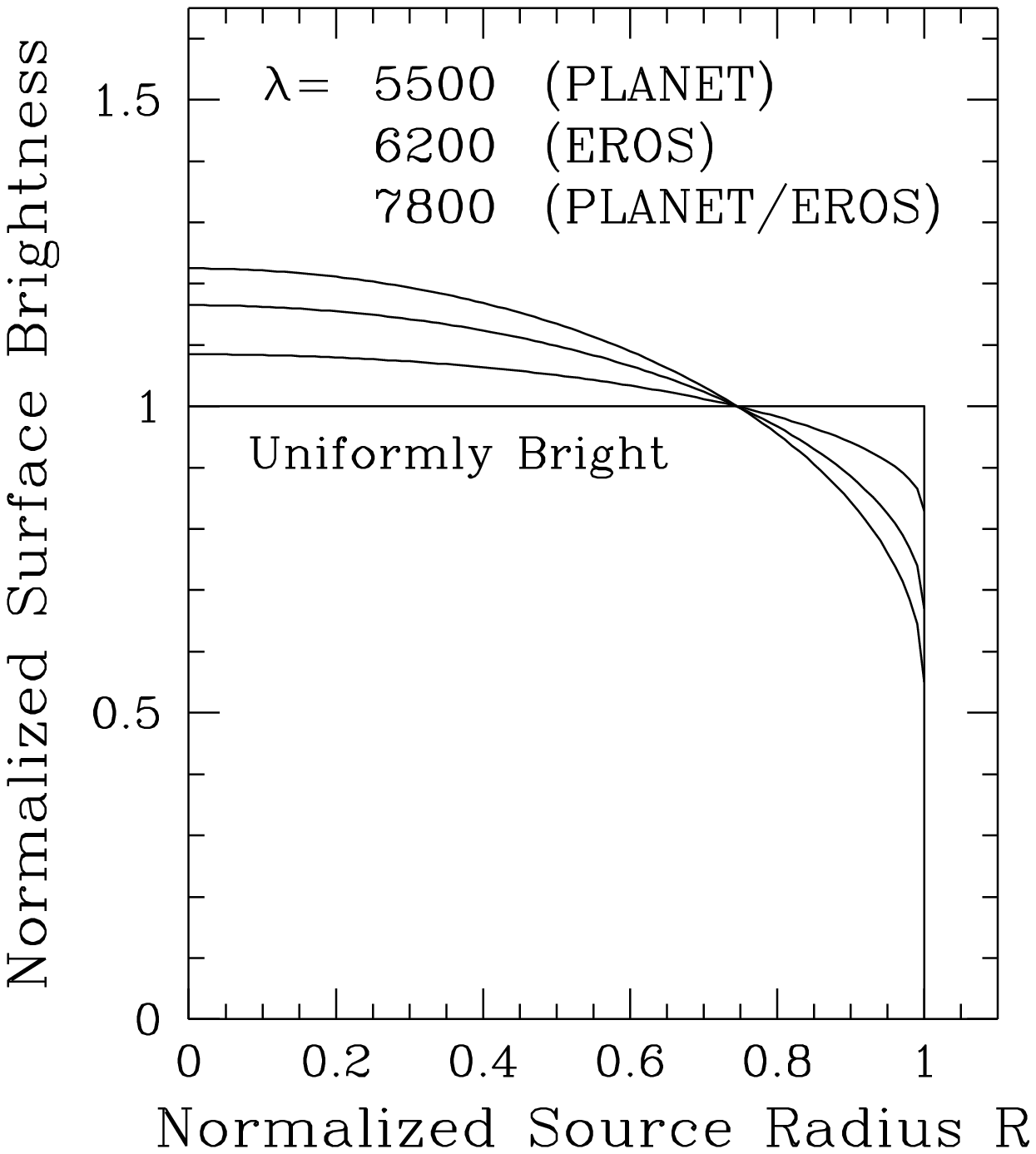}{4.0cm}{0}{35}{35}{5}{27}
\vglue -3.2cm
\caption{{\bf Left:~} The binary lens event MACHO~98-SMC-01 was exceptionally 
well-sampled due to the efforts of five microlensing teams. 
{\bf Right:}  Data over the second fold caustic crossing allowed 
one-parameter profiles to be deduced at several wavelengths 
for this A-dwarf source star in the SMC. (Based on Afonso et al. 2000). 
} 
\end{figure}

A third microlensing limb-darkening measurement was made for the cool (4750~K) 
giant source star in the Galactic bulge event MACHO~97-BLG-41.  This was a 
cusp-crossing event in which rotation of a binary lens was measured 
for the first time (Albrow et al. 2000, see also Menzies et al. 2000).  
The linear parameter $\Gamma_{I} = 0.42 \pm 0.09$, corresponding 
to $c_{I} = 0.52 \pm 0.10$, determined from the light curve agreed well with that 
of $c_{I} \approx 0.56$ from atmospheric models (Claret et al. 1995) 
of stars of the appropriate temperature and gravity.

The photometric precision required to recover linear limb-darkening 
coefficients with 10\% accuracy has been estimated recently by 
Rhie \& Bennett (2000), and found to be about 1\% (relative photometry). 
As inspection of the residuals in Fig.~2 clearly reveals, a worldwide 
network of 1m-class telescopes is quite capable of this precision.  
At the current rate of observed caustic crossing events, the community 
can thus expect that microlensing will provide 2 or 3 limb-darkening measurements 
per year; indeed more results are now in preparation.
 
\subsection{Spectroscopy}

\begin{figure}
\vglue -0.3cm
\plotfiddle{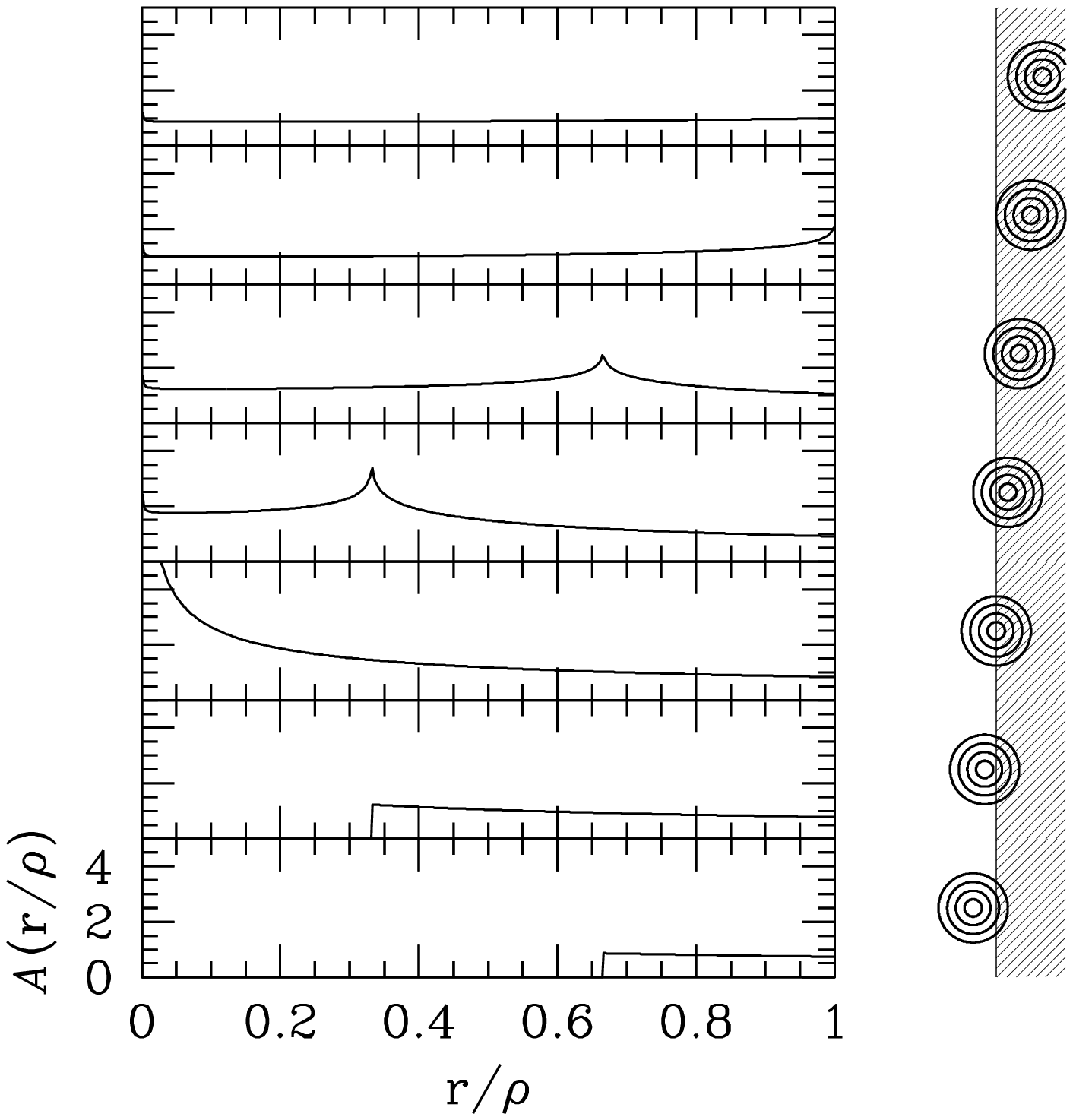}{3.9cm}{0}{35}{35}{-180}{-100}
\plotfiddle{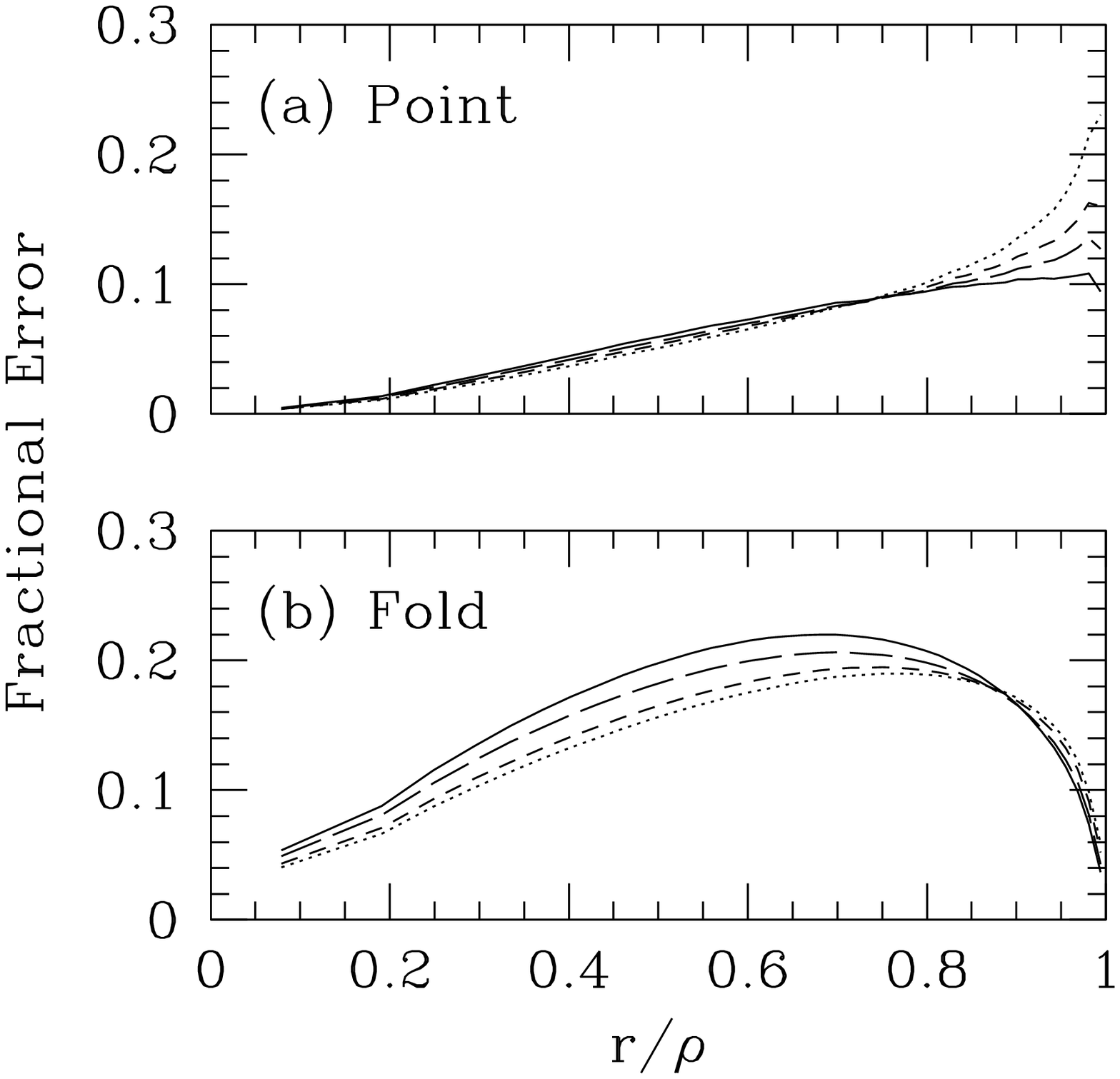}{3.9cm}{0}{30}{30}{0}{32}
\vglue -3.1cm
\caption{{\bf Left:~} The azimuthally-averaged magnification as a function of 
fractional position $r/\rho$ on stellar disk as a source (with $\rho = 1$) 
moves (from top to bottom) out of a caustic over a fold singularity.
{\bf Right:} Fractional error in the recovered intensity profile ($\delta I/I$) 
as a function of $r/\rho$ for a direct point transit (top) 
and a fold crossing (bottom).  Typical results are displayed for  
a 2m telescope, including limb-darkening effects for $VIK$ passbands 
(Gaudi \& Gould 1999). 
} 
\end{figure}

The magnification boost provided by microlensing can yield 
higher S/N spectra of faint sources than 
would otherwise be possible.  Lennon et al. (1996) used 
microlensing to measure the effective temperature, gravity and 
metallicity of a G-dwarf in the bulge; at the time of the caustic 
boost, the 3.5m NTT had the collecting power of a 17.5m aperture.
In another case, lithium was detected in a bulge turn-off 
star using Keck and microlensing (Minitti et al. 1998).

Attempts have been made to perform time-resolved spectroscopy during 
caustic crossings in order to detect the varying spectral signatures
expected (Loeb \& Sasselov 1995; Valls-Gabaud 1998) 
as light from different positions across the stellar disk 
(and thus different optical and physical depths) is differentially 
magnified.  The caustic alert provided by the 
MACHO team (Alcock et al. 2000) allowed Lennon and colleagues (1996, 1997) to 
take spectra over the peak of the fold crossing in MACHO~96-BLG-3, 
though these did not extend far enough down the decline to detect 
spectral differences.  Temporal coverage was also insufficient to detect 
strong spectral changes during the point caustic transit in MACHO~95-BLG-30,  
although slight equivalent width variations in TiO (Alcock et al. 1997) 
and H$\alpha$ (Sasselov 1998) lines may have been seen.

\section{Stellar Tomography: A New Era of Stellar Atmosphere Physics}

\begin{figure}
\vskip -0.2cm
\plotfiddle{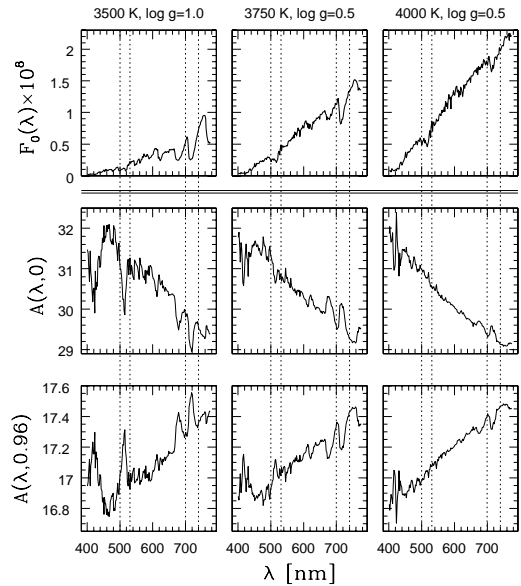}{4cm}{0}{40}{40}{-120}{-165}
\vglue 3.2cm
\caption{Unlensed (top) and microlensed spectra for three different 
model atmospheres of a cool giant.  Microlensed spectra are for a point 
caustic at the center (middle) and limb (bottom) of the source.  
Vertical dotted lines mark TiO bands (Heyrovsk\'y, Sasselov \& Loeb 2000). 
} 
\end{figure}

The ability of current microlensing collaborations to predict (fold) caustic crossings 
$\sim$1-3 days in advance opens a new era for stellar atmosphere physics.    
Follow-up teams, separately or in collaboration with existing microlensing networks, 
could obtain spectrophotometric data on auxiliary telescopes 
in order to take advantage of the magnification and resolution afforded by  
microlens caustics.  

Theoretical expectations of how microlensing may contribute to our understanding 
of stellar atmospheres over the next decade are encouraging.  
Gaudi \& Gould (1999) have 
simulated a 2m-class telescope with a 1\AA-resolution spectrograph 
continuously observing a typical $V = 17$ ($\rho = 0.02$) bulge source  
undergoing a fold caustic crossing of 7 hours duration.  They find 
that the intensity profile can be recovered to a precision of 10-20\% 
using a spatial resolution of 10\% across the star for most wavelengths. 
Most of the spatial resolution generated by fold caustics comes from the 
period in which the trailing limb is exiting the caustic curve.  
Direct transits of point caustics could provide even more reliable 
intensity profiles (Fig.~5), but due to geometric factors are much more 
unlikely to occur.  These results appear to agree 
with the estimates derived from different approaches 
(Hendry et al. 1998; Gray \& Coleman 2000).

Loeb \& Sasselov (1995) noted that the ``light curves'' of 
atmospheric emission lines that are most prominent in 
the cool outer layers of giants will experience sharp peaks when the 
limb (rather than the center) of the source crosses a caustic.  
Recently, this work has been extended (Heyrovsk\'y, Sasselov 
\& Loeb 2000) to lines across the whole optical spectrum to show 
how time-resolved spectroscopy 
during a caustic crossing can discriminate between different atmospheric 
models.  For a given impact parameter, a particular line may appear in 
emission or absorption depending on the temperature structure of the star 
(Fig.~6).  Since the duration of the caustic crossing is limited, 
4m and 8m telescopes will be required to obtain the highest spectral resolution.

Experience and modeling thus indicates that any aperture can perform 
microlensing tomography of stellar atmospheres in the next decade, 
{\it provided that the community is willing to 
reschedule telescope access on a few days notice.\/}

\newpage

\subsection*{Acknowledgements}

The author gratefully acknowledges support from 
the Nederlandse Organisatie voor Wetenschappelijk Onderzoek 
(GBE 614-21-009) and the meeting LOC.

\end{document}